%% file: rand.tex
\documentclass[letterpaper,twocolumn,10pt]{article}
\usepackage{usenix2019_v3}

\usepackage[binary-units]{siunitx}
\DeclareSIUnit{\billion}{\text{billion}}

\usepackage{tikz}
\usetikzlibrary{arrows, calc, positioning, decorations.markings, decorations.pathreplacing}
\usepackage{nicefrac}

\usepackage{pgfplots}                                                           
\usepackage{pgf}                                                           
\usepgfplotslibrary{statistics}
\usepackage{csvsimple}

\usepackage{adjustbox}
\pgfplotsset{compat=1.13}

\usepackage{blindtext} 

\usepackage{csquotes} 

\usepackage{xspace}
\usepackage{xcolor}
\usepackage{xifthen}
\newcommand{\ifequals}[3]{\ifthenelse{\equal{#1}{#2}}{#3}{}}

\definecolor{darkred}{rgb}{0.831, 0, 0.063}
\definecolor{darkgreen}{rgb}{0.043, 0.768, 0.003}
\definecolor{darkblue}{rgb}{0, 0.2, 0.6}

\usepackage{amssymb}  

\newtoggle{blinded}
\togglefalse{blinded}

\iftoggle{blinded}{
\newcommand{\cve}{\emph{CVE-2020-blinded}\xspace}
}
{
\newcommand{\cve}{\emph{CVE-2020-6616}\xspace} %
}

\usepackage{subcaption}
\usepackage{listings} 

\lstdefinelanguage{ASM}{
    morekeywords={b, ble, blt, bne, bx, bl, ldr, str, push, pop, mov, add, sub},
    keywordstyle=\color{blue},
    sensitive=false, 
    morecomment=[l]{//}, 
    morecomment=[s]{/*}{*/}, 
    morestring=[b]", 
} %
\lstdefinelanguage{none}{
  identifierstyle=
}

\usepackage[nolist]{acronym}
\begin{acronym}
\acro{SDR}{Software-Defined Radio}
\acro{AGC}{Automatic Gain Control}
\acro{GCI}{Global Coexistence Interface}
\acro{ECI}{Enhanced Coexistence Interface}
\acro{AFH}{Adaptive Frequency Hopping}
\acro{FEM}{Front-End Module}
\acro{SP3T}{Single Pole, Triple Throw}
\acro{HCI}{Host Controller Interface}
\acro{A2DP}{Advanced Audio Distribution Profile}
\acro{SCO}{Synchronous Connection-Oriented}
\acro{GPIO}{General Purpose Input Output}
\acro{ASLR}{Address Space Layout Randomization}
\acro{LMP}{Link Management Protocol}
\acro{LCP}{Link Control Protocol}
\acro{LNA}{Low-Noise Amplifier}
\acro{EIR}{Extended Inquiry Response}
\acro{CRC}{Cyclic Redundancy Check}
\acro{RTOS}{Real-Time Operating System}
\acro{QEMU}{Quick Emulator}
\acro{UART}{Universal Asynchronous Receiver Transmitter}
\acro{MITM}{Machine-in-the-Middle}
\acro{MAC}{Media Access Control}
\acro{BLE}{Bluetooth Low Energy}
\acro{ACL}{Asynchronous Connection-Less}
\acro{DSP}{Digital Signal Processing}
\acro{SDIO}{Secure Digital Input Output}

\acro{BCS}{Bluetooth Core Scheduler}
\acro{ELF}{Executable and Linking Format}
\acro{GIAC}{Global Inquiry Access Code}
\acro{JSON}{JavaScript Object Notation}
\acro{L2CAP}{Logical Link Control and Adaptation Protocol}
\acro{LMP}{Link Management Protocol}
\acro{PTM}{Pseudo Terminal Master}
\acro{PTS}{Pseudo Terminal Slave}
\acro{QEMU}{Quick Emulator}
\acro{RCE}{Remote Code Execution}
\acro{RFU}{Reserved for Future Use}
\acro{SVC}{Supervisor Call}
\acro{UART}{Universal Asynchronous Receiver Transmitter}
\acro{WICED}{Wireless Internet Connectivity for Embedded Devices}
\acro{MMIO}{Memory Mapped Input/Output}
\acro{LE}{Bluetooth Low Energy}
\acro{IoT}{Internet of Things}
\acro{IDE}{Integrated Development Environment}
\acro{ARM}{Advanced RISC Machine}
\acro{LM}{Link Manager}

\acro{RTOS}{Real-Time Operating System}
\acro{DMA}{Direct Memory Access}
\acro{RXDMA}{Receive Direct Memory Access}
\acro{NVRAM}{Non-Volatile Random-Access Memory}
\acro{ROM}{Read-Only Memory}
\acro{Rx}{Receive}
\acro{Tx}{Transmit}
\acro{SPI}{Serial Peripheral Interface}
\acro{MSP}{Main Stack Pointer}
\acro{PSP}{Process Stack Pointer}
\acro{RF}{Radio Frequency}
\acro{BLOB}{Binary Large Object}
\acro{SCO}{Synchronous Connection Oriented}
\acro{UAF}{Use-After-Free}
\acro{FHS}{Frequency Hop Sync}
\acro{CRC}{Cyclic Redundancy Check}
\acro{PDU}{Protocol Data Unit}
\acro{NFC}{Near Field Communication}
\acro{JPEG}{Joint Photographic Experts Group}
\acro{LPE}{Local Privilege Escalation}
\acro{DEP}{Data Execution Prevention}
\acro{XN}{eXecute Never}
\acro{LCP}{Link Control Protocol}
\acro{GATT}{Generic Attribute}
\acro{PoC}{Proof of Concept}
\acro{EDR}{Enhanced Data Rate}
\acro{MWS}{Mobile Wireless Standards}
\acro{HAL}{Hardware Abstraction Layer}
\acro{LR}{Link Register}
\acro{eSIM}{embedded-SIM}
\acro{RSP}{Remote SIM Provisioning}
\acro{PC}{Program Counter}
\acro{MD}{More Data}
\acro{LLID}{Logical Link Identifier}
\acro{SFI}{Serial Flash Interface}
\acro{EEPROM}{Electrically Erasable Programmable Read-Only Memory}
\acro{TOFU}{Trust On First Use}
\acro{CVE}{Common Vulnerabilities and Exposure}
\acro{PCIe}{Peripheral Component Interconnect Express}
\acro{SECI}{Serial Enhanced Coexistence Interface}
\acro{HID}{Human Interface Device}
\acro{DoS}{Denial of Service}

\acro{PRNG}{Pseudo-Random Number Generator}
\acro{RNG}{Random Number Generator}
\acro{HRNG}{Hardware Random Number Generator}
\acro{SSP}{Secure Simple Pairing}
\acro{CRC}{Cyclic Redundancy Check}
\acro{NDA}{Non-Disclosure Agreement}
\acro{SoC}{System on Chip}
\acro{ECDH}{Elliptic-curve Diffie--Hellman}
\acro{KDF}{Key Derivation Function}
\acro{SMP}{Security Manager Protocol}
\acro{CRC}{Cyclic Redundancy Check}

\end{acronym}

\usepackage[colorinlistoftodos,prependcaption,disable]{todonotes} 
\presetkeys%
    {todonotes}%
    {inline,backgroundcolor=orange}{}

\begin{document}

\date{}

\title{\Large \bf Firmware Insider: Bluetooth Randomness is Mostly Random}
\iftoggle{blinded}{
\author{{Authors blinded for review.}}
}
{
\author{
{\rm J{\"o}rn Tillmanns}\\
Secure Mobile Networking Lab\\
TU Darmstadt
\and
{\rm Jiska Classen}\\
Secure Mobile Networking Lab\\
TU Darmstadt
\and
{\rm Felix Rohrbach}\\
\hspace*{3.7em}Cryptoplexity\hspace*{3.7em}\\
TU Darmstadt
\and
{\rm Matthias Hollick}\\
Secure Mobile Networking Lab\\
TU Darmstadt
} 
}

\maketitle

\begin{abstract}
Bluetooth chips must include a \ac{RNG}. This \ac{RNG} is used internally within
cryptographic primitives but also exposed to the operating system for chip-external
applications.
In general, it is a black box with security-critical authentication
and encryption mechanisms depending on it.
In this paper, we evaluate the quality of \acp{RNG} in various \emph{Broadcom} and
\emph{Cypress} Bluetooth chips. We find that the \ac{RNG} implementation significantly changed
over the last decade. Moreover, most devices implement an insecure \ac{PRNG} fallback.
Multiple popular devices, such as the \emph{Samsung Galaxy S8} and its variants as well as an \emph{iPhone}, rely on the weak fallback due to
missing a \ac{HRNG}. We statistically evaluate the output of various \acp{HRNG} in chips
used by hundreds of millions of devices.
While the  \emph{Broadcom} and \emph{Cypress} \acp{HRNG} pass advanced tests, it remains indistinguishable for users if a
Bluetooth chip implements a secure \ac{RNG} without an extensive analysis as in this paper.
We describe our measurement methods and publish our tools to enable further public testing.
\end{abstract}

\input{sections/intro.tex}
\input{sections/background.tex}

\input{sections/firmware.tex}

\input{sections/hrng.tex}
\input{sections/discussion.tex}

\input{sections/conclusion.tex}


\section*{Acknowledgments}

We thank \emph{Apple}, \emph{BlueKitchen}, \emph{Broadcom}, \emph{Cypress}, \emph{Google}, \emph{Qualcomm}, and \emph{Samsung} for handling the responsible disclosure requests.

\iftoggle{blinded}{
Further acknowledgments are blinded for review.
}
{
We also thank Jakob Link for enabling remote access to the \emph{Samsung Galaxy S8} for further testing.
Moreover, we thank Maximilian Tschirschnitz for pointing out a detail within the Bluetooth specification.
Furthermore, we thank Matthias Ringwald and Teal Starsong for the paper feedback.

This work has been funded by the German Federal
Ministry of Education and Research and the Hessen State Ministry for
Higher Education, Research and the Arts within their joint support of
the National Research Center for Applied Cybersecurity ATHENE, as well as by the Deutsche Forschungsgemeinschaft (DFG) -- SFB 1119 -- 236615297.
}


\section*{Availability}
The \ac{RNG} measurement assembler patches and scripts for the \emph{Broadcom} and \emph{Cypress}
Bluetooth chips are openly available on \emph{GitHub}.
They are hosted as examples for \emph{InternalBlue} on \url{https://github.com/seemoo-lab/internalblue}.


\bibliographystyle{plain}
\bibliography{bibliographies}

\end{document}

%% file: sections/intro.tex

\section{Introduction}

High-quality random numbers ensure security within cryptographic methods.
The Bluetooth 5.2 specification makes no exception to this---it requires a Bluetooth chip to provide at least a \ac{PRNG} compliant to \emph{FIPS PUB 140-2}~\cite[p.~953]{bt52,fips1402}. Various security-relevant functions depend on random numbers, such as generating authentication and encryption keys, nonces within \ac{SSP}, or passkeys used in authentication. Moreover, embedded and \ac{IoT} devices with \ac{BLE} can request random numbers locally from the chip~\cite[p.~2521]{bt52}. Thus, a Bluetooth chip's \ac{RNG} quality can also be relevant to external applications.

According to the specification, the \ac{RNG} shall be tested against \emph{FIPS SP800-22}~\cite{fipssp80022}. Recommended test suites are \emph{Diehard}, \emph{Dieharder}, and the \emph{NIST tools}~\cite{dieharder,nisttest}.
We choose the \emph{Dieharder} suite, as it implements the most extensive tests.
Verification of the randomness properties is not straightforward. Even though these test suites exist, the local \ac{HCI} that enables the operating system to request random numbers from the chip introduces a lot of overhead per request and only returns an \SI{8}{\byte} number. In contrast, the \emph{Dieharder} test suite requires at least \SI{1}{\giga\byte} of unique data to return meaningful results, making randomness tests a challenge.

Statistical tests might not uncover all issues within an \ac{RNG}.
The underlying implementation can be a \ac{HRNG} or \ac{PRNG}. The latter is part of the chip's firmware and could have flaws only detectable by reverse-engineering the implementation.
While the \ac{HRNG} could also have flaws that stay undetected by statistical tests, this goes beyond the analysis of this paper.
The according datasheets, which are only available for a few of the older chips, do not cover any details about the \ac{HRNG}.

In this paper, we reverse-engineer and measure the \ac{RNG} implementations of \emph{Broadcom} and \emph{Cypress} Bluetooth chips\footnote{\emph{Cypress} acquired parts of \emph{Broadcom} in 2016~\cite{cypressbroadcom}, and the code bases diverged since then.}.
In particular, we analyze the \acp{RNG} on \num{20} chips, including those on the most recent \emph{iPhones} and \emph{Samsung Galaxy S} series.
To this end, we root or jailbreak those devices, dump their firmware with the \emph{InternalBlue} framework~\cite{mantz2019internalblue}, and reverse-engineer these.
Moreover, we write custom patches that increase the \ac{RNG} output to make measurements with the \emph{Dieharder} test suite feasible.
These patches need to be customized for each chip and the according operating systems, which are \emph{Android}, \emph{Linux}, \emph{iOS}, and \emph{macOS}. 
Our main findings are as follows:

\todo{if time allows: measure the HRNG speed}
\todo{if time allows: btstack le rand usage?}

\begin{itemize}
\item All chips (except the \emph{Samsung Galaxy S8} variants and an \emph{iPhone}) contain a  \ac{HRNG}. Yet, most implement a \ac{PRNG} fallback in case the \ac{HRNG} is not available.
\item The \ac{PRNG} is based on various predictable inputs, which significantly reduces its entropy. We show that an active over-the-air attacker can infer and manipulate \ac{PRNG} values.
\item Multiple patches were applied to the \ac{RNG} implementation over time, such as adding a cache. The \ac{PRNG} fallback is no longer present in the most recent \emph{Broadcom} chips but \emph{Cypress} still maintains this code.
\item An \ac{HRNG} is missing on the European version of the \emph{Samsung Galaxy S8} and its variants in the \emph{S8+} and \emph{Note 8}, which was sold approximately \SI{40}{million} times.
\item After responsibly disclosing the issue, the firmware was patched in May 2020 in \emph{Samsung}'s \emph{Android} release as well as in \emph{iOS 13.5} for an unspecified \emph{iPhone}.
\end{itemize}

Flaws in the \ac{RNG} allow attacks on Bluetooth authentication and encryption mechanisms.
Recently, attacks on the \ac{ECDH} key exchange and key negotiation showed how sensitive these mechanisms are~\cite{2018:biham, knob, antonioli20bias, confusion}.
While these flaws were in the specification itself, \ac{RNG} issues are implementation-specific and rather opaque.

We initiated responsible disclosure with \emph{Broadcom}, \emph{Cypress}, and a selection of their customers on January 12, 2020.
The weak \ac{PRNG} implementation was assigned \cve.
\emph{Broadcom} denied that the \ac{PRNG} fallback was used on any of their devices despite it being present in their firmware---until we found and reported that the \ac{HRNG} code and register mappings were missing on the \emph{Samsung Galaxy S8} on February 1, 2020.

This paper is structured as follows:
In \autoref{sec:background}, we provide background information how the \ac{RNG} is used within security-critical parts of the Bluetooth specification.
We reverse-engineer 20 firmware variants in \autoref{sec:firmware}
and continue with \ac{HRNG} and \ac{PRNG} measurements as well as \ac{PRNG} attacks in \autoref{sec:hardware}.
We discuss further aspects in \autoref{sec:discussion}.
We conclude our findings in \autoref{sec:conclusion}.

\begin{figure*}[!t]

\center

	\begin{center}
	\begin{tikzpicture}[minimum height=0.55cm, scale=0.8, every node/.style={scale=0.8}, node distance=0.7cm]
	
    \node[inner sep=0pt] (iphonea) at (5,0.5)
    {\includegraphics[height=1.5cm]{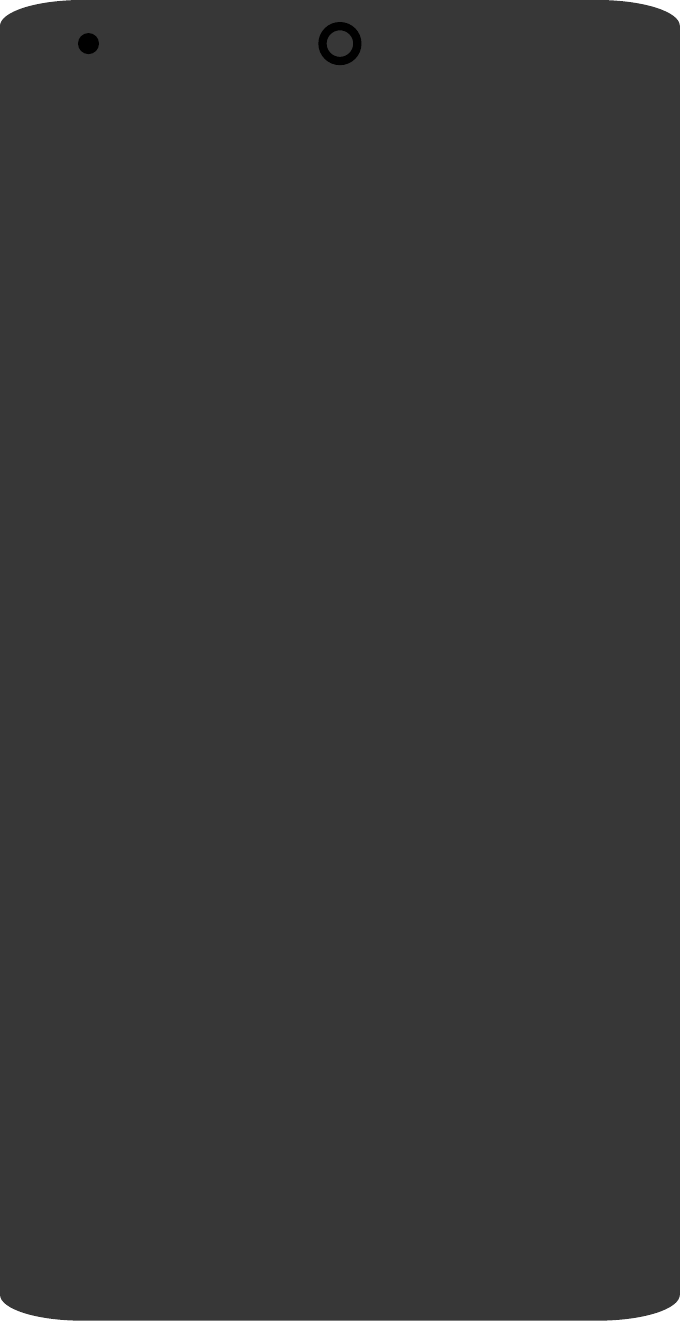}}; 
    \node[inner sep=0pt] (nexusx) at (5,0.5)
    {\includegraphics[height=1.3cm]{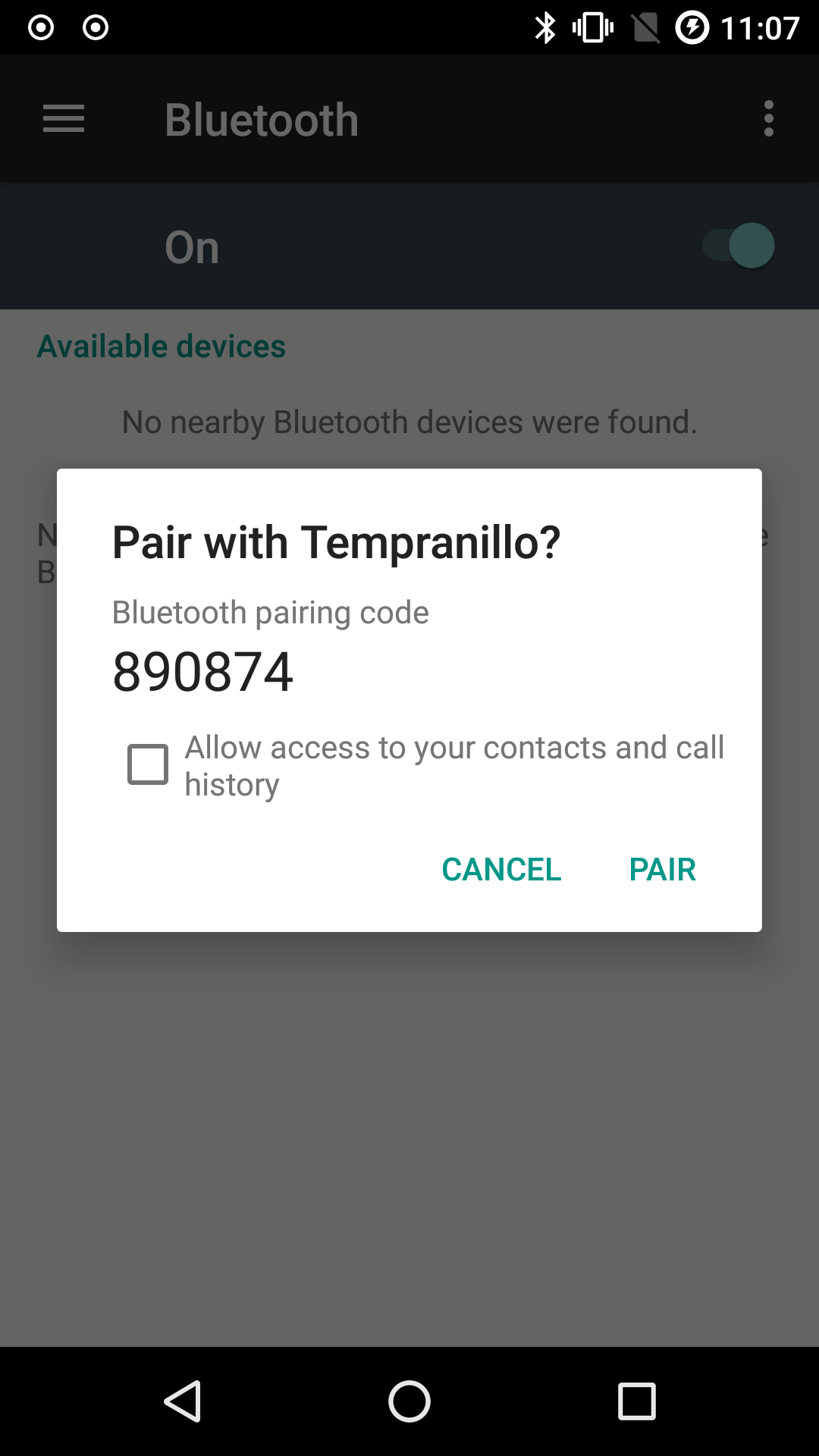}};  
    \node[inner sep=0pt] (iphone) at (16,0.5)
    {\includegraphics[height=1.5cm]{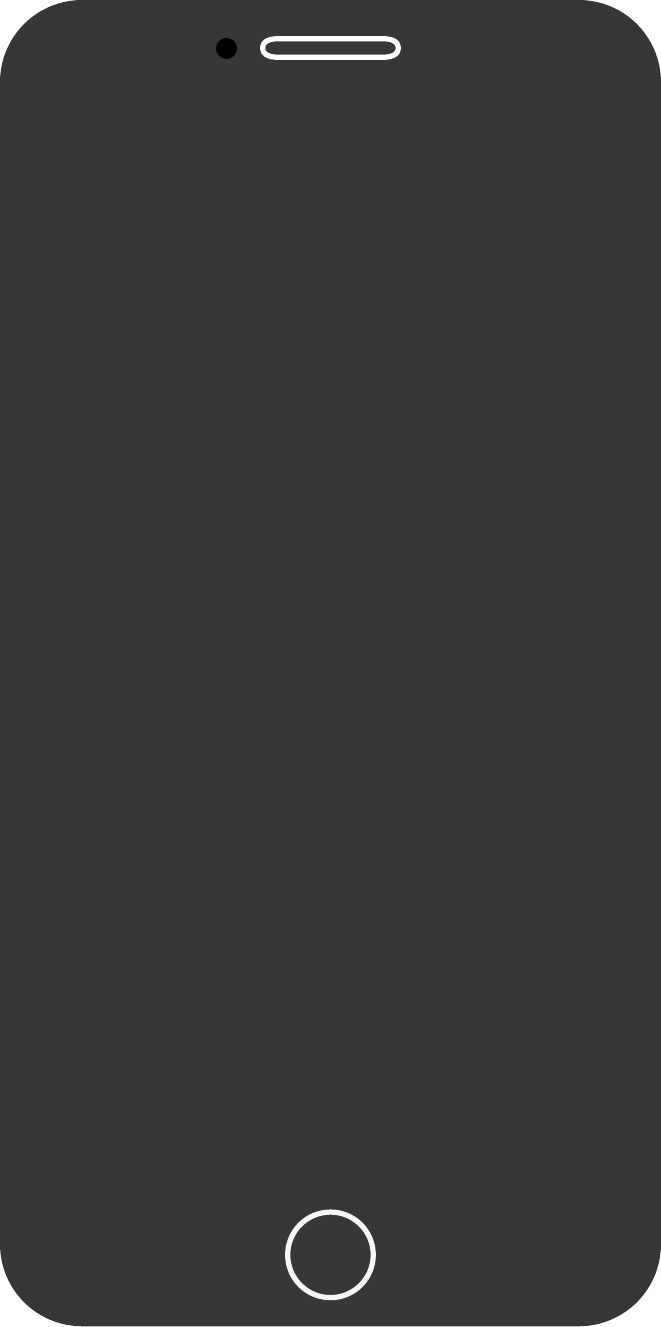}}; 
    \node[inner sep=0pt] (iphonex) at (16,0.5)
    {\includegraphics[height=1.2cm]{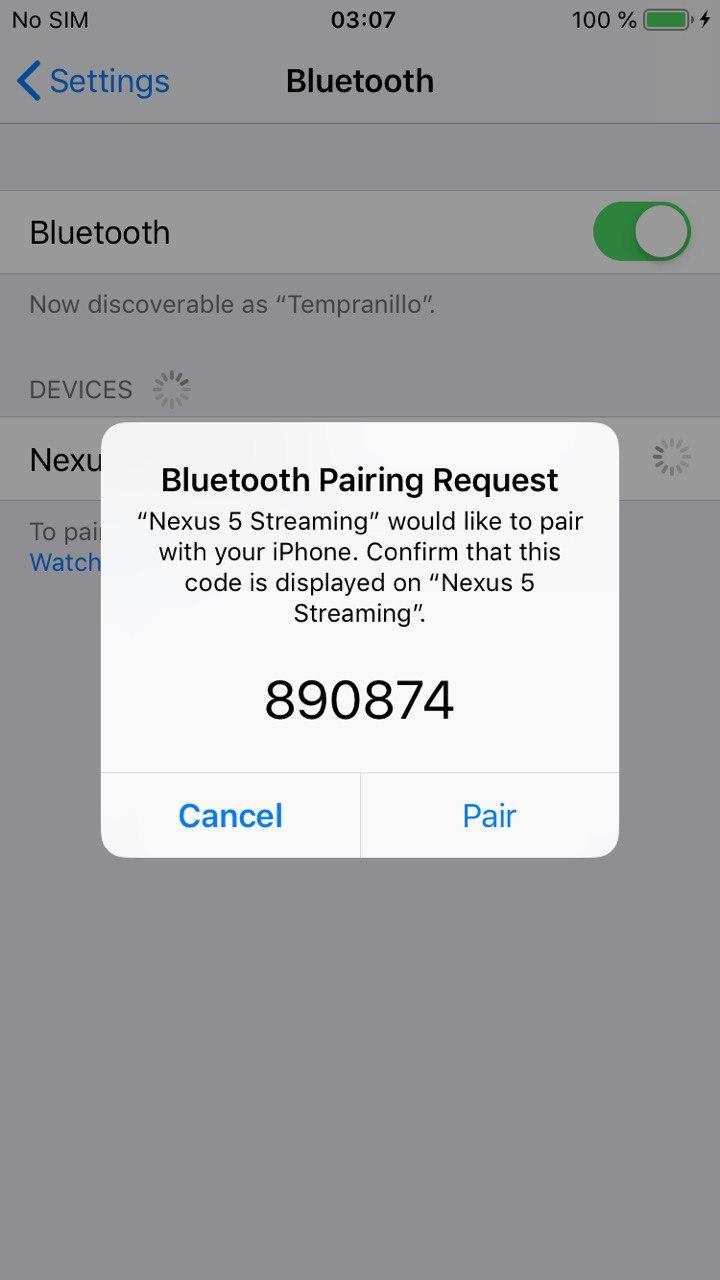}};

    \node[inner sep=0pt] (attack) at (13,0.5)
    {\includegraphics[height=1.5cm]{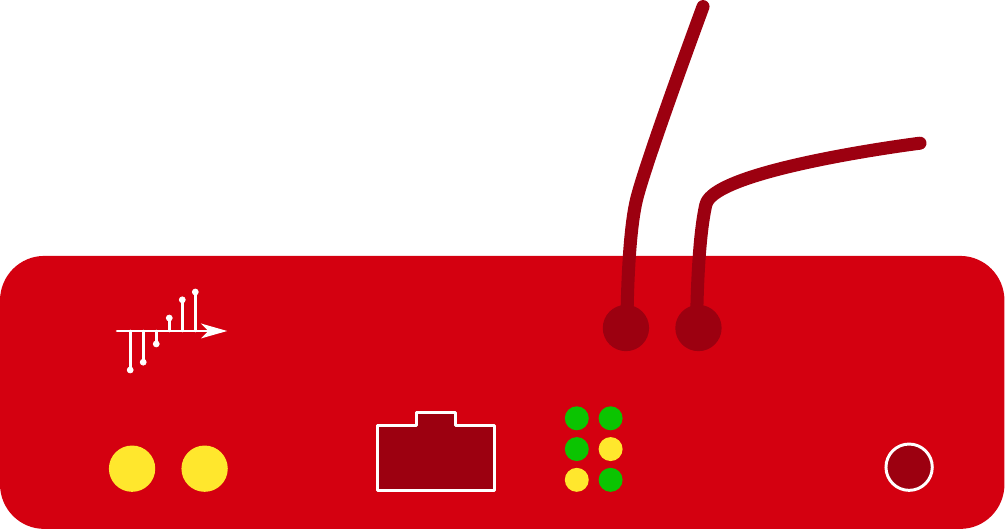}};

    \node[below=of iphone.south, anchor=south, yshift=0.3cm] (iphonetxt) {Device $B$};
    \node[below=of iphonea.south, anchor=south, yshift=0.3cm] (iphonetxt) {Device $A$};
    \node[below=of attack.south, anchor=south, yshift=0.3cm] (attacktxt) {MITM $E$};

    \draw[-,gray,dashed,thick] (5,-0.8) -- (5,-10);
    \draw[-,gray,dashed,thick] (16,-0.8) -- (16,-10);
    \draw[-,darkred!30,dashed,thick] (13,-0.8) -- (13,-7.25);
    
    \node[align=right,left] at (4.5, -2.5) {Select $Na$ randomly \\ \\ Set $ra$ and $rb$ to \num{0} \\ \\ \\}; 
    \node[align=left,right] at (16.5, -2.5) {\textcolor{darkred}{Select $Nb$ randomly} \\ \\ Set $rb$ and $ra$ to \num{0} \\ \\ Compute commitment: \\ $Cb = f1(PKbx, $\textcolor{darkred}{$PKex$}$, Nb, 0)$}; 
    \node[align=right,left] at (13, -2.5)  {\textcolor{darkred}{Guess $Nb$}}; 

    \path[<-] (13.2,-3.5) edge node[sloped, anchor=center, above, text width=0.5cm,yshift=-0em] {$Cb$} (15.8,-3.5);
    \path[-,color=darkred] (13.3,-3.7) edge node[sloped, anchor=center, above, text width=0.5cm,yshift=-0em] {} (14,-3.3);
    
    \node[align=right,left] at (13, -4)  {\textcolor{darkred}{Check if $Cb==f1(PKbx, PKex, Nb, 0)$} \\ \textcolor{darkred}{$Cb'=f1(PKex, PKax, Nb, 0)$ }};

    \path[<-,color=darkred] (5.2,-5) edge node[sloped, anchor=left, above, text width=5cm,yshift=-0em] {\textcolor{darkred}{$Cb'$}} (12.8,-5);
    
    \path[->] (5.2,-5.75) edge node[sloped, anchor=left, above, text width=5cm,yshift=-0em] {$Na$} (12.8,-5.75);
    \path[-,color=darkred] (5.3,-5.95) edge node[sloped, anchor=center, above, text width=0.5cm,yshift=-0em] {} (6,-5.55);
    
    \node[align=right,left] at (13, -6.5)  {\textcolor{darkred}{Choose $Na'$ s.t. $g(PKax, PKex, Na, Nb)$} \\ \textcolor{darkred}{$==g(PKex, PKbx, Na', Nb)$ }};

    \path[->,color=darkred] (13.2,-6.75) edge node[sloped, anchor=center, above, text width=0.5cm,yshift=-0em] {\textcolor{darkred}{$Na'$}} (15.8,-6.75);
    
    \path[<-] (5.2,-7.5) edge node[sloped, anchor=left, above, text width=8cm,yshift=-0em] {$Nb$} (15.8,-7.5);
     
    \node[align=right,left] at (4.5, -8) {Check if \textcolor{darkred}{$Cb'$}$==f1($\textcolor{darkred}{$PKex$}$, PKax, Nb, 0)$\\ Abort if check fails!}; 
     
    \node[align=right,left] at (4.5, -9)  {$Va=g(PKax, $\textcolor{darkred}{$PKex$}$, Na, Nb)$}; 
    \node[align=left,right] at (16.5, -9) {$Vb=g($\textcolor{darkred}{$PKex$}$, PKbx, $\textcolor{darkred}{$Na'$}$, Nb)$};

	\draw[draw=darkblue,thick,fill=darkblue!10](5.2,-8) rectangle (15.8, -10);
    \node[align=left,right, text width=9.8cm] at (5.5, -9)  {$Va$ and $Vb$ are 6-digit numbers displayed on both devices. The user confirms with \emph{yes} or \emph{no}. \\ \emph{Just Works} mode is the same as \emph{Numeric Comparison} but without displaying a number, thus, it does not have active MITM protection. };

	\end{tikzpicture}
	\end{center}
	
\caption{\emph{Numeric Comparison} protocol attack vector~\cite[p. 986]{bt52}.}
\label{fig:numericcomp}
\end{figure*}
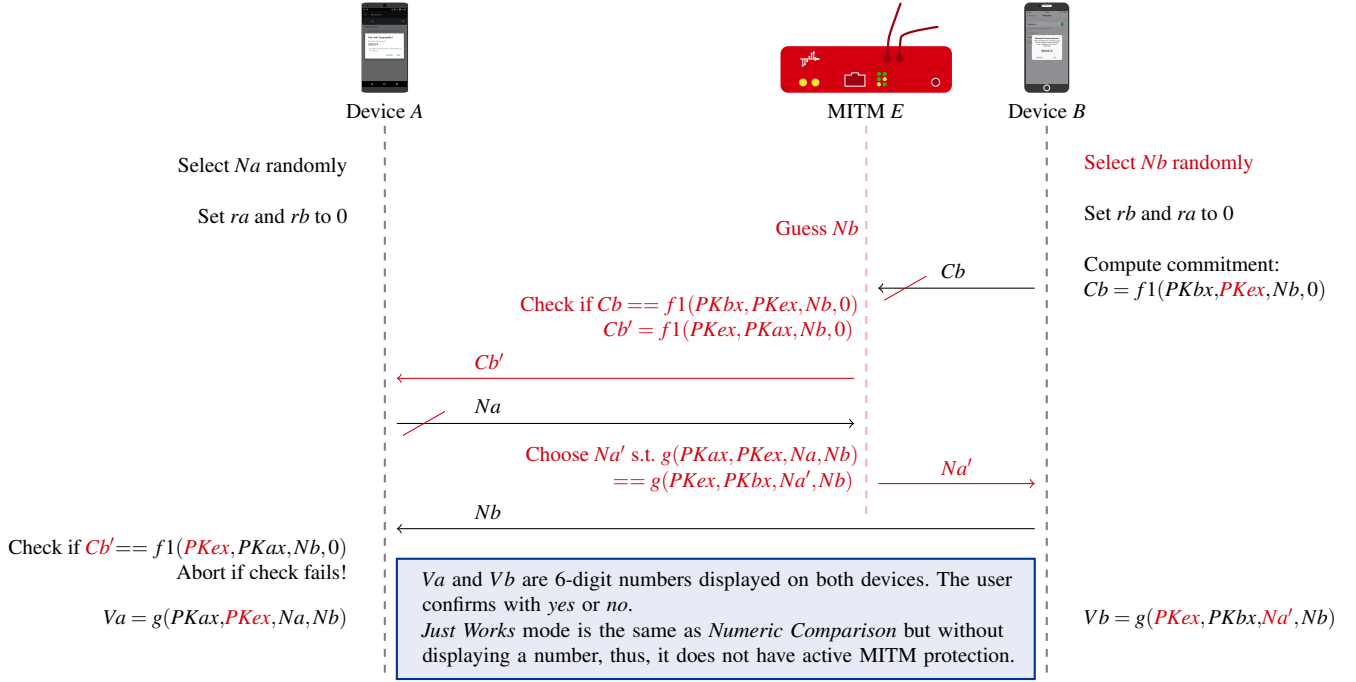

%% file: sections/background.tex

\section{RNG Usage Within Bluetooth}
\label{sec:background}

A predictable or known \ac{RNG} has a severe impact on security-relevant functions in Bluetooth.
The Bluetooth 5.2 specification only vaguely mentions that this is the case~\cite[p.~953]{bt52}, but
does not provide any context which functions break if the \ac{RNG} does not meet the requirements.
Thus, we outline concrete examples where the \ac{RNG} matters. These examples are not a complete list---this would exceed the scope of this paper.
In general, random numbers are used in many more places within the Bluetooth specification, and
an unknown number of applications within host applications.

\subsection{Active \acs{MITM} Attack on Pairing with Numeric Comparison}
\label{ssec:numericComparison}

During the initial pairing of two Bluetooth devices, the protocol requires user input to prevent an active \ac{MITM} attack, as no previous
key material exists on any of the devices to identify the other one. Bluetooth \ac{SSP} provides different methods for this kind of authentication,
a commonly used one is \emph{Numeric Comparison}~\cite[p. 985]{bt52}: Both devices display a 6-digit number to the user and the user checks whether both
numbers are equal. While \emph{Numeric Comparison} was introduced for Classic Bluetooth with \ac{SSP}, it is also supported by \ac{BLE}
under the name \emph{LE Secure Connections} since Bluetooth 4.2~\cite[p. 269]{bt52}, because the previous \emph{LE Legacy Pairing} is broken
by design~\cite{ryan2013bluetooth}.
\todo{Teal says: LE in LE Secure Connections not defined. It's a name, though?}

\emph{Numeric Comparison} happens in the second stage of authentication, after both parties have exchanged their \ac{ECDH} public keys.
Device $B$, the
non-initiating party, starts by generating a random number $Nb$ and sending a commitment $Cb$ of $Nb$ and both public keys to device $A$. Device $A$ 
answers by generating a random number, $Na$, as well, which it sends to $B$. Finally, $B$ sends $Nb$ to $A$, $A$ checks whether $Cb$ is indeed a commitment
of $Nb$ and both parties hash $Na$, $Nb$ and both public keys to a 6-digit number, which is then displayed on both devices.

If there was an active \ac{MITM} attacker, the public keys that are fed into the hash function would differ and the two devices would 
display different values with high probability. However, this depends on the attacker not being able to change either $Na$ or $Nb$ to search
for a second pre-image of the hash function, as the 6-digit output does not protect against brute-force attacks. Therefore, the protocol
is designed in a way that no party can decide on their random number after seeing the random number of the other party---$A$ by sending its number first
and $B$ by committing to its randomness before seeing $Na$.

The security of the \emph{Numeric Comparison} protocol crucially relies on the hiding property of the commitment, i.e., that no party can calculate $Nb$ from
the commitment $Cb$. As the commitment is deterministic, this requires $Nb$ to be a number with high entropy. Now, if an attacker can predict $Nb$ 
or small set of possible $Nb$s, they can check for $Nb$ by calculating the commitment again and comparing it to $Cb$. If successful, they can
use their knowledge of $Nb$ to create a value $Na^\prime$, sent to $B$, such that both devices display the same 6-digit code even though the public
keys fed into the hash function differ.
For the full attack, see \autoref{fig:numericcomp}.

\subsection{LE Randomness Within Android}
\label{ssec:androidle}

\ac{BLE} uses the \ac{SMP} for pairing in \emph{LE Secure Connections} and \emph{LE Legacy Pairing}~\cite[p. 1666]{bt52}. \ac{SMP} resides on top of \ac{LE} data \acp{PDU} and initiates secure keys.
Encryption is then started and stopped with \ac{LE} link control \acp{PDU} independent from \ac{SMP}.

When initiating a \ac{BLE} pairing on \emph{Android 6--10}, the \ac{HCI} command \path{LE_Rand} is called multiple times.
This way, \emph{Android} receives specification-compliant random numbers within its Bluetooth stack, but keeps control over the keys itself.
We further investigate the current \emph{Android} \texttt{master} branch as of April 2020~\cite{androidbt}. All randomness-related function calls originate from the file \path{smp_keys.cc}.

During a \ac{BLE} pairing in \emph{Just Works} mode, the functions \path{smp_create_private_key} and \path{smp_start_nonce_generation} are called. These use random inputs from the Bluetooth chip's \ac{RNG}.
\path{smp_create_private_key} directly fills the \ac{ECDH} private key by calling the Bluetooth \ac{RNG} via \ac{HCI} four times in a row and then calculates a public key based on this.
Thus, an attacker who knows the internal \ac{RNG} state can infer the \ac{ECDH} private key.

After creating the \ac{ECDH} key pair, the value generated with \path{smp_start_nonce_generation} is sent in plaintext over-the-air and also contains \ac{RNG} data.
Then, an \ac{LE} link control \ac{PDU} \path{LL_ENC_REQ} is sent~\cite[p. 2898]{bt52}, which also contains values from the Bluetooth \ac{RNG}, but is generated locally on the chip without \emph{Android} interaction.

This means that on \emph{Android}, directly after using the Bluetooth \ac{RNG} for creating a private key, further \ac{RNG} values that could leak the internal \ac{RNG} state are sent in plaintext over-the-air.
Note that this behavior is not required by the Bluetooth specification, e.g., \emph{iOS 13}
does not call the Bluetooth \ac{RNG} upon \ac{SMP} key generation.

\todo{BLE p. 3029, encryption start procedure, states the hard requirement of a compliant rng for the iv and skd (session key diversifier).

p. 1679 explains things but wtf this is a big mess mixing le secure connections and le legacy pairing all over -.-

the security manager specification for LE is on p. 1618ff. le legacy pairing is broken anyway, so only le secure connections is relevant.
le secure connections: p. 1699 shows diagram. le uses l2cap that's why it's not in lcp except the encryption start req.}

%% file: sections/firmware.tex
\begin{table*}[!t]
\renewcommand{\arraystretch}{1.3}
\caption{\ac{RNG} implementation variants in 20 \emph{Broadcom} and \emph{Cypress} chips.}
\label{tab:fwvariants}
\centering
\scriptsize
\begin{tabular}{|c|l|l|r|r|l|l|}
\hline
\textbf{Variant} & \textbf{Chip} & \textbf{Device} & \textbf{Build Date} & \textbf{HRNG Location} & \textbf{PRNG} & \textbf{Cache} \\
\hline
1 & BCM2046A2 & iMac Late 2009 & 2007 & 0xE9A00, 3 regs & Minimal (inline) & No \\
 & BCM2070B0 & MacBook 2011 & Jul 9 2008  & 0xE9A00, 3 regs &  &  \\
 & BCM20702A1 & Asus USB Dongle, Thinkpad T420 & Feb (?) 2010  &  0xEA204, 3 regs &  &  \\
\hline
2 & BCM4335C0 & Google Nexus 5 & Dec 11 2012 &  0x314004, 3 regs & Advanced (inline) & No \\
 & BCM4345B0 & iPhone 6 & Jul 15 2013 & 0x314004, 3 regs &  &  \\
 & BCM20703A1	 & MacBook Pro early 2015 & Dec 23 2013 &  0x314004, 3 regs &  &  \\ 
 & BCM43430A1 & Raspberry Pi 3/Zero W & Jun 2 2014 &  0x352600, 3 regs &  &  \\
 & BCM4345C0 & Raspberry Pi 3+/4 & Aug 19 2014 & 0x314004, 3 regs &  &  \\
 & BCM4358A3 & Samsung Galaxy S6, Nexus 6P & Oct 23 2014 &  0x314004, 3 regs &  &  \\
 & BCM4345C1 & iPhone SE & Jan 27 2015 & 0x314004, 3 regs &  &  \\
 & BCM4364B0 & MacBook/iMac 2017--2019 & Aug 21 2015  & 0x352600, 3 regs &  &  \\
 & BCM4355C0 & iPhone 7 & Sep 14 2015  & 0x352600, 3 regs &  &  \\
 & BCM20703A2 & MacBook/iMac 2016--2017 & Oct 22 2015  & 0x314004, 3 regs &  &  \\
 & CYW20719B1 & Evaluation board & Jan 17 2017 &  0x352600, 3 regs &  &  \\
\hline
3 & CYW20735B1 & Evaluation board & Jan 18 2018 &  0x352600, 3 regs & Advanced, 8 regs & Yes, breaks after 32 elements \\
 & CYW20819A1 & Evaluation board & May 22 2018 &  0x352600, 3 regs & Advanced, 5 regs & Yes, with minor fixes \\
\hline
4 & BCM4347B0 & Samsung Galaxy S8/S8+/Note 8 & Jun 3 2016 & \textcolor{darkred}{Not mapped} & \textcolor{darkred}{Only option} & No \\ 
\hline
5 & BCM4347B1 & iPhone 8/X/XR & Oct 11 2016 & 0x352600, 4 regs & None & Asynchronous 32x cache \\
 & BCM4375B1 & Samsung Galaxy S10/S20 & Apr 13 2018 & 0x352600, 4 regs &  &  \\
 & BCM4378B1 & iPhone 11/SE2 & Oct 25 2018  & 0x602600, 4 regs & &  \\
\hline
\end{tabular}
\end{table*}

\section{Firmware Variant Analysis}
\label{sec:firmware}

In the following, we compare the \ac{RNG} function across 20 firmware versions of even more devices.
Sales numbers are rather vague and older devices might no longer be in use, but in total, the chips we analyzed
are used in hundreds of millions if not even in a billion of devices.
An overview of these is shown in \autoref{tab:fwvariants}. Since the firmware is located
in the ROM and built during device development, the firmware build date is at least a year
before the device release date. For some \emph{MacBooks} and the \emph{Raspberry Pi 4},
it is even five years. Moreover, some build dates do not represent the exact state of the 
libraries that were compiled into it. Thus, the table is sorted by firmware variants.

We obtain firmware dumps and locate the \ac{RNG} function as described in \autoref{ssec:fwsymbols}.
Based on this analysis, we identify five variants (see \autoref{ssec:fwvariants}).
We provide a pseudo-code description of the main variants in \autoref{ssec:fwimpl}.

\subsection{Firmware Symbols and Comparison Methods}
\label{ssec:fwsymbols}
\emph{InternalBlue} enables firmware ROM and RAM dumps on various operating systems to extract the firmware
from \emph{Broadcom} and \emph{Cypress} chips~\cite{mantz2019internalblue}. The extraction requires physical device access and rooting or jailbreaking of mobile devices. Moreover, such firmware dumps do not contain any symbols or strings. Thus, reverse-engineering the \ac{RNG} is not straightforward.

The \emph{Cypress} evaluation kits allow to develop an \ac{IoT} application with \emph{WICED Studio}~\cite{wiced}.
This application is running on the same ARM core as the Bluetooth firmware.
Thus, \emph{WICED Studio} includes a file called \texttt{patch.elf} to link the application during the
build process. This patch file contains global function and variable names. Furthermore, each board's \path{*map*.h} file
contains hardware register names.
\emph{WICED Studio} also includes symbols in an \ac{ARM} compiler specific format for the \emph{Broadcom} \emph{BCM20703A2} Wi-Fi/Bluetooth combo chip in the files \path{ram_ext.symdefs} and \path{20703mapa0.h}.

We use these partial symbols for static firmware analysis of the \ac{RNG}, which is internally provided by the function \path{rbg_rand}.
Since the \ac{HRNG} is only mapped to two different locations throughout all variants, we can locate these in further firmware
variants.
The oldest variant is mapped to previously unknown addresses, but uses the same magic value \texttt{0x200FFFFF} when accessing the \ac{HRNG}.

\subsection{Identified Variants}
\label{ssec:fwvariants}
\autoref{tab:fwvariants} compares the implementations of the function \path{rbg_rand}.
Regardless of the implementation variant, \path{rbg_rand} always returns a \SI{4}{\byte} random number.
Overall, we identify five different variants within firmware released over more than a decade.

\paragraph{1. Minimal PRNG Fallback}
The oldest variant contains the worst \ac{PRNG} fallback. If no \ac{HRNG} is available, it skips waiting for a new random number and performs a static calculation based on the current time---which might lead to zero entropy.

\paragraph{2. Advanced PRNG Fallback}
This variant is similar to variant \emph{(1)} but the \ac{PRNG} implementation is more advanced and does not only consider time.

\paragraph{3. Cache and Advanced PRNG Fallback}
\emph{Cypress} introduced a cache that is constantly filled with \num{32} \SI{4}{\byte} random numbers, probably to increase performance. Within this variant, also the registers used by the \ac{PRNG} vary and some that tend to be static were removed.

\paragraph{4. Advanced PRNG Only}
This variant is similar to variant \emph{(2)} but has the \ac{PRNG} as \emph{only} option. The \ac{HRNG} access is missing within the firmware and dynamic analysis reveals that it is not mapped at all.

\paragraph{5. Asynchronous Cache and No PRNG}
The newest variant is a complete rewrite of the \texttt{rbg} library by \emph{Broadcom}. The cache is filled asynchronously in the background, the \ac{HRNG} has an additional register, and the \ac{PRNG} fallback was removed.

\begin{figure}[b!]
\tikzstyle{myarrow}=[->, >=open triangle 90, thick]
\tikzstyle{func_node}=[rectangle, draw=black, rounded corners, fill=white,	text centered, anchor=north, text=black, font=\ttfamily]

\begin{tikzpicture}
	\draw[draw=darkblue,thick,fill=darkblue!10](0.27,-0.45) rectangle (7.5, -1.6);
	\draw[draw=darkblue,thick,fill=darkblue!10](0.55,-2.1) rectangle (7.5,-3.55);
	\draw[draw=darkblue,thick,fill=darkblue!10](0.27,-4.35) rectangle (7.5,-5);

\draw (0,0) node[inner sep=0,anchor=north west] {
	\begin{lstlisting}[language=C, label=lst:rbg_rand_mac,caption={Variant 1--4: \texttt{rbg\_rand} on the \emph{BCM20703A2 MacBook}.}]
uint32 rbg_rand(void){
	uint32 RBG_CONST_READY = 0x200FFFFF;
	uint32 RBG_CONST_STATUS_OK = 0xFFFFF000; 
	uint32 _rbg_ret_val;
	uint32 *prng_store;

	if (*rbg_status << 12 == RBG_CONST_STATUS_OK) { 
		while (*rbg_status != RBG_CONST_READY) {
			wdog_restart();
			*hrng_control = 1;
		}
		_rbg_ret_val = *rbg_random_num;
	} else {
		// rbg_get_psrng (inline)
	}
	*prng_store = _rbg_ret_val;
	return (uint32) _rbg_ret_val;
}
	\end{lstlisting}};
	
	\node[func_node, text width= 1.3cm, text height=0.7em, xshift=7.8cm, yshift = -0.2cm](init) {\scriptsize init};
	\node[func_node, text width= 1.3cm, text height=0.7em, yshift = -0.15cm, below=of init](get_rbg) {\scriptsize get\_rbg};	
	\node[func_node, text width= 1.3cm, text height=0.7em, yshift = -0.75cm, below=of get_rbg](teardown) {\scriptsize teardown};
\end{tikzpicture}
\end{figure}

\begin{figure}[b!]
\begin{lstlisting}[language=C,caption={Variant 2--4: Advanced PRNG fallback on the \emph{BCM20703A2 MacBook}.}, label=lst:rbg_get_psrng_mac]
bcs_waitForBTclock()
data_array[0] = *@\textbf{dc\_nbtc\_clk}@    // Bluetooth clock
data_array[1] = *@\textbf{timer1value}@    // System clock
data_array[2] = *@\textbf{dc\_fhout}@
data_array[3] = *@\textbf{agcStatus}@ 
data_array[4] = *@\textbf{rxInitAngle}@
data_array[5] = *@\textbf{spurFreqErr1}@
data_array[6] = *@\textbf{rsPskPHErr5}@

if (*rgb_psrng_control == 0){
	__rt_memcpy(data_array[7], mm_top, 4)
	len = 0x2c
} else {
	len = 0x20
	data_array[7] = *@\textbf{psrng\_store}@
}
_rbg_ret_val = @\textbf{crc32\_update}@(0xFFFFFFFF, data_array, len)
bcs_releaseBTclock()
*rgb_psrng_control	= 1	
\end{lstlisting}
\end{figure}


\begin{figure}[b!]
\begin{lstlisting}[language=C,caption={Variant 1: Minimal PRNG fallback on the \emph{BCM2046A2 iMac}.}, label=lst:rbg_get_psrng_imac]
return clock ^ ((16 * static_register + 180) << 20) ^ static_value[4 * static_register]
\end{lstlisting}
\end{figure}

\subsection{Implementation Details}
\label{ssec:fwimpl}

Variant \emph{(2)}, the \emph{Advanced \ac{PRNG} Fallback}, is the most common variant that we found.
Since we have partial symbols for this variant, we can reconstruct the original logic of the \ac{HRNG} access (\autoref{lst:rbg_rand_mac}) and \ac{PRNG} implementation (\autoref{lst:rbg_get_psrng_mac}).
Note that except from the complete rewrite in variant \emph{(5)}, the code throughout all variants is very similar.

The \ac{HRNG} is accessed with the three mapped registers \path{rbg_control}, \path{rbg_status}, and \path{rbg_random_num}.
A new random number is requested by writing \texttt{1} to \path{rbg_control}. The \path{rbg_status} indicates if the \ac{HRNG} is available in general and if it currently has a fresh random number.
The random number itself is accessible via \path{rbg_random_num}. If the \ac{HRNG} is available in general, the \path{rbg_rand} function enters an endless loop until a new random value is available.
As this loop might take longer, it pets the watchdog.

The \ac{PRNG} depends on two timing values, which are the Bluetooth clock \path{dc_nbtc_clk} and system clock \path{timer1value}.
The \SI{4}{\byte} Bluetooth clock is comparably slow, runs in steps of \SI{312.5}{\micro\second}~\cite[p. 415]{bt52}, and is shared over-the-air. In comparison, the system clock passes faster
and is not directly known. Depending on the platform, it is either \SI{2}{\byte} or \SI{4}{\byte}.
The remaining \ac{PRNG} values are acquired from signal reception characteristics. If the \ac{PRNG} was not used before, it is initialized based on the current memory contents. Otherwise,
the last value within the \ac{PRNG} store is taken. Since the \path{rbg_rand} function only returns \SI{4}{\byte}, the entropy of these values is combined by calculating a \ac{CRC}.

The older variant \emph{(1)}, the \emph{Minimal \ac{PRNG} Fallback}, is even worse. The fallback only performs a static calculation based on the current time, as shown in \autoref{lst:rbg_get_psrng_imac}.
The \SI{4}{\byte} time register is increased by one every \SI{0.005}{\second}. 

%% file: sections/hrng.tex

\begin{table}[!b]
\renewcommand{\arraystretch}{1.3}
\caption{\ac{HRNG} test results.}
\label{tab:hrng}
\centering
\scriptsize
\begin{tabular}{|l|l|c|c|}
\hline
\textbf{Chip} & \textbf{Device} & \textbf{Samples} & \textbf{Test} \\
\hline
BCM4335C0 & Google Nexus 5 & \SI{2.7}{\giga\byte} &  \\
BCM43430A1 & Raspberry Pi 3/Zero W & \SI{1.3}{\giga\byte} & Dieharder \\
BCM4345B0 & iPhone 6 & \SI{1.8}{\giga\byte} & passed \\
BCM4355C0 & iPhone 7 & \SI{1.0}{\giga\byte} &  \\
BCM4345C0 & Raspberry Pi 3+/4 & \SI{1.4}{\giga\byte} &  \\
BCM4358A3 & Samsung Galaxy S6, Nexus 6P & \SI{2.1}{\giga\byte} &  \\
CYW20719B1 & Evaluation board & \SI{1.4}{\giga\byte} &  \\
CYW20735B1 & Evaluation board & \SI{1.6}{\giga\byte} &  \\
CYW20819A1 & Evaluation board & \SI{1.2}{\giga\byte} &  \\
\hline
BCM2046A2 & iMac Late 2009 & --- & \checkmark ~HRNG \\
BCM20703A1 & MacBook Pro early 2015	 &  &  \\
BCM4375B1 & Samsung Galaxy S10/S20 &  &  \\
BCM4347B1 & iPhone 8/X/XR	 &  & \\
BCM4378B1 & iPhone 11/SE2 &  & \\
\hline
\end{tabular}
\end{table}

\input{pics/s8_dc_nbtc_clk.tex}

\section{HRNG and PRNG Tests}
\label{sec:hardware}

The function \path{rbg_rand} can be analyzed statically based on firmware dumps,
however, it must also be validated on physical hardware.
For example, the code accessing the \ac{HRNG} was missing in the firmware of the \emph{Samsung Galaxy S8}.
However, the code might be present and the test if the \ac{HRNG} register is available could still return \texttt{false}.
Furthermore, if the \ac{PRNG} is accessed in absence of the \ac{HRNG}, the entropy of the registers that it is accessing
becomes relevant.
In the following, we first measure the \ac{HRNG} in \autoref{ssec:hrngmes} and then measure the \ac{PRNG} fallback for the devices using it in \autoref{ssec:prngmes}.
Additionally, we discuss specification-compliant attacks to change \ac{PRNG} inputs and infer the current \ac{PRNG} state over-the-air in \autoref{ssec:prngattack}.

\subsection{HRNG Measurements}
\label{ssec:hrngmes}
The only existing interface to acquire random numbers of the Bluetooth chip is the \acf{HCI} command \path{LE_Rand}~\cite[p. 2521]{bt52}.
After successful execution, this command is answered with an \ac{HCI} event containing an \SI{8}{\byte} random number. In theory, this could
already be used on any off-the-shelf device to measure the \ac{RNG} quality. However, access via \path{LE_Rand} has two drawbacks.
First, the \emph{Dieharder} test suite requires at least \SI{1}{\giga\byte} of random data to successfully complete all tests.
Second, further functions within the Bluetooth firmware might access the \ac{RNG} in parallel and influence the results.

Utilizing \emph{InternalBlue}~\cite{mantz2019internalblue}, we make the following modifications to \emph{Broadcom} and \emph{Cypress}
chips to collect \ac{HRNG} data.
We substitute the original \path{rbg_rand} routine with \texttt{return 0} to gain exclusive \ac{HRNG} access.
Except from the evaluation boards, all chips are Wi-Fi/Bluetooth combo chips. Thus, we also disable Wi-Fi to exclude further side-effects.
Then, we collect random data in chunks. The chunk size depends on the memory available on the specific device, but on most devices, we found free chunks of \SI{20.480}{\kilo\byte}.
This allows to collect \num{4096} measurements of \SI{5}{\byte} values.
After each \SI{4}{\byte} returned by the \ac{HRNG}, we insert one static check byte, which we remove before running \emph{Dieharder}. This ensures that no other process was using this memory area in parallel and is faster to insert than a checksum.
Once collection of a chunk is finished, we issue an asynchronous \ac{HCI} event to notify the host. Then, the host collects the random data using the vendor-specific command \path{Read_RAM}, which can read \SI{251}{\byte} chunks.

The \emph{Dieharder} test suite usually expects an endless source of randomness, such as \path{/dev/urandom} on UNIX-like operating systems.
In contrast, the Bluetooth \ac{HRNG} results are stored in files. As a baseline, we run \emph{Dieharder} on files produced with \path{/dev/urandom}
as input and find that at least \SI{1}{\giga\byte} of random data is required to pass all tests. Therefore, this is the lower bound of data collected
for each chip. With the modifications listed above, data collection takes approximately one day, depending on the chip's interface
and operating system. \autoref{tab:hrng} shows how much data was collected per chip and that the \acp{HRNG} in all chips passed the \emph{Dieharder} tests.

Each data extraction for the \emph{Dieharder} test suite still takes comparably long and requires custom patches.  However, given that the extensively tested \acp{HRNG} passed all
tests, already checking if a \ac{HRNG} is present on a chip is an important information. \emph{InternalBlue} implements everything
required for such a basic check out-of-the-box---reading chip memory and sending \ac{HCI} commands. The hardware registers listed in \autoref{tab:fwvariants} must contain one
 \SI{4}{\byte} random number. This number is indeed the one used by the \path{rbg_rand} function
if it changes with actions like pairing. On devices supporting \ac{BLE}, the \ac{HRNG} can be triggered with the \ac{HCI} command \path{LE_Rand}.
Firmware versions implementing a cache are required to get a call to \path{LE_Rand} multiple times before the \ac{HRNG} is used.

The second half of \autoref{tab:hrng} contains devices from which we could not collect samples but that indeed have an \ac{HRNG}.
These devices include the \emph{iMac Late 2009}, the newest \emph{Samsung Galaxy S} series, and the \emph{iPhone 11} and \emph{SE2}.
On the \emph{iMac Late 2009}, we were not able to program custom patches because RAM is very limited on that chip.
Moreover, its interface is slow~\footnote{One specification-compliant \ac{HCI} randomness event contains \SI{8}{\byte}, while our events fill the maximum possible event payload of \SI{251}{\byte}. The timing of \ac{HCI} events is almost constant independent on their size. Assuming that our \SI{251}{\byte} data extraction runs one day, the \SI{8}{\byte} data extraction would require a month per device. The exact \ac{HCI} speed depends on the \emph{InternalBlue} and host-specific implementation of each device.}, making it unrealistic to extract \SI{1}{\giga\byte} of data.
The \emph{Samsung Galaxy S10} and \emph{S20} series use the same chip. However, \emph{Broadcom} improved firmware
security and added stack canaries, which prevent calling functions out of context without additional modifications within our patches.
As of now, there is no \emph{InternalBlue} support for \ac{PCIe} Bluetooth chips as on the
most recent \emph{iPhone 11} and \emph{SE2}. Nonetheless, the chip can be tested with \emph{BlueTool} on
jailbroken \emph{iPhones}. \emph{BlueTool} is an \emph{Apple}-internal tool included on \emph{iOS}.
It is utilized for
driver initialization, but it also has a rudimentary command-line interface with \ac{HCI} support.



\begin{table}[!b]
\renewcommand{\arraystretch}{1.3}
\caption{\ac{PRNG} inputs on the \emph{Samsung Galaxy S8}.}
\label{tab:prng}
\centering
\scriptsize
\begin{tabular}{|c|l|p{4.7cm}|}
\hline
\textbf{Address} & \textbf{Register} & \textbf{Entropy} \\
\hline
--- & \path{Rand} & \textcolor{darkred}{Previous \SI{4}{\byte} random value (leaks over-the-air)}\\
\texttt{0x318088} & \path{dc_nbtc_clk} & \textcolor{darkred}{Bluetooth clock, publicly available over-the-air}\\
\texttt{0x32A004} & \path{timer1value} & \textcolor{darkred}{Hardware clock, \SI{4}{\byte} ``random'' before first leak, \newline unpatched attacks for clock reset available}\\
\texttt{0x3186A0} & \path{dc_fhout} & \textcolor{orange!70!red}{Changes a bit (\texttt{0x02}--\texttt{0x50})}\\
\texttt{0x410434} & \path{agcStatus} & \textcolor{orange!70!red}{Changes a bit (\texttt{0xc00} during whole measurement, slight changes within \texttt{0xc$nn$} after reboot)} \\
\texttt{0x41079C} & \path{rxInitAngle} & \textcolor{orange!70!red}{Changes a bit but within similar range}\\
\texttt{0x4100AC} & \path{spurFreqErr1} & \textcolor{darkred}{Constant \SI{2}{\byte} value (\texttt{0x04ed}, also after reboot)}\\
\texttt{0x410548} & \path{rxPskPhErr5} & \textcolor{darkred}{Always 0}\\
\hline
\end{tabular}
\end{table}

\subsection{PRNG Measurements}
\label{ssec:prngmes}

The \emph{Samsung Galaxy S8} series and an \emph{iPhone} are missing an \ac{HRNG}.
As there is a \ac{PRNG} fallback, the firmware is still able to generate
somewhat random numbers instead of being obviously broken for an outsider without firmware knowledge.
Most likely the \ac{RNG} test required to pass the specification did not include a firmware review,
and, thus, only the checksumed combination of the registers the \ac{PRNG} uses was measured.

The \ac{PRNG} is accessing multiple hardware registers, listed in \autoref{tab:prng}.
While we do not have any symbols for the \emph{Samsung Galaxy S8} firmware, it is using the same registers as
the \emph{BC20703A2 MacBook} chip, and also the same code as previously shown in \autoref{lst:rbg_get_psrng_mac}.

\begin{figure}[!b]
\begin{tikzpicture}[every node/.style={font=\footnotesize}]
\begin{axis}[ybar stacked, bar width=1.6pt, minor y tick num = 3, xtick={2,10,...,80},height=5cm, width=0.52\textwidth,xmin=-1,xmax=83]
\addplot[style = {darkblue, fill=darkblue!30}] coordinates {
        (2, 6062) (3, 6224) (4, 5904) (5, 6142) (6, 8595) (7, 5808) 
        (8, 8238) (9, 7545) (10, 5901) (11, 3703) (12, 4186) (13, 6167) 
        (14, 8289) (15, 5971) (16, 5879) (17, 4170) (18, 6064) (19, 7789) 
        (20, 6480) (21, 8095) (22, 6314) (23, 8146) (24, 8134) (25, 11726) 
        (26, 6146) (27, 9854) (28, 6160) (29, 3956) (30, 8003) (31, 4096) 
        (32, 6177) (33, 6947) (34, 3963) (35, 5974) (36, 4200) (37, 8381) 
        (38, 10626) (39, 8250) (40, 4096) (41, 6111) (42, 6083) (43, 6063) 
        (44, 6290) (45, 6786) (46, 4236) (47, 6062) (48, 8020) (49, 5242) 
        (50, 4489) (52, 6317) (54, 6225) (56, 6226) (58, 6121) (60, 4022) 
        (62, 5502) (64, 5825) (66, 6226) (68, 6387) (70, 6384) (72, 4006) 
        (74, 8325) (76, 6205) (78, 3992) (80, 10094) 
};
\end{axis}
\end{tikzpicture}
\caption{Histogram of values of \texttt{dc\_fhout} observed on a \emph{Samsung Galaxy S8}.}
\label{fig:dc_fhout}
\end{figure}

\input{pics/s8_rxInitAngle.tex} 

We measure the \ac{PRNG} registers with similar firmware modifications as for the \ac{HRNG} measurements and also disable \mbox{Wi-Fi}.
Over \num{1000} rounds, we store \num{4096} \SI{4}{\byte} values to the chip's internal RAM. Then, we collect them via \ac{HCI}.
Due to the current \emph{InternalBlue} implementation on \emph{Android 9}, \ac{HCI} introduces a delay, which is \SI{3.075}{\second} on average in our measurements.
Thus, data about each register is collected over a time span of approximately \SI{51}{\minute}.
Each measurement round takes roughly \SI{2.58}{\milli\second} on the chip itself.

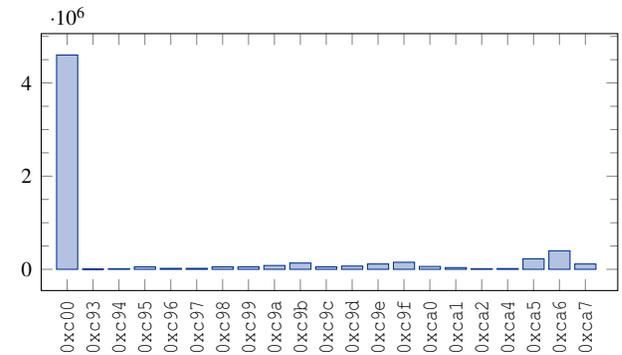
\begin{figure}[!b]
\begin{tikzpicture}[every node/.style={font=\footnotesize}]
\begin{axis}[ybar stacked, bar width=8pt, minor y tick num = 3, height=5cm, width=0.52\textwidth, 
xticklabel style={rotate=90},
xticklabels={\texttt{0xc00}, \texttt{0xc93}, \texttt{0xc94}, \texttt{0xc95}, \texttt{0xc96}, \texttt{0xc97}, \texttt{0xc98}, \texttt{0xc99}, \texttt{0xc9a}, \texttt{0xc9b}, \texttt{0xc9c}, \texttt{0xc9d}, \texttt{0xc9e}, \texttt{0xc9f}, \texttt{0xca0}, \texttt{0xca1}, \texttt{0xca2}, \texttt{0xca4}, \texttt{0xca5}, \texttt{0xca6}, \texttt{0xca7}},
xtick={0, 1, 2, 3, 4, 5, 6, 7, 8, 9, 10, 11, 12, 13, 14, 15, 16, 17, 18, 19, 20},
xmin=-1, xmax=21.25]

\addplot[style = {darkblue, fill=darkblue!30}] coordinates {
        (0, 4599808) (1, 4096) (2, 12288) (3, 53248) (4, 20480) (5, 20480) 
        (6, 53248) (7, 53248) (8, 81920) (9, 135168) (10, 53248) (11, 69632) 
        (12, 114688) (13, 151552) (14, 61440) (15, 36864) (16, 12288) (17, 16384) 
        (18, 225280) (19, 397312) (20, 114688) 
};

\end{axis}
\end{tikzpicture}
\caption{Histogram of values of \texttt{agcStatus} observed on a \emph{Samsung Galaxy S8}, two measurements, first one was constant \texttt{0xc00}.}
\label{fig:agcStatus}
\end{figure}

During the whole experiment, \path{spurFreqErr1} and \path{rxPskPhErr5} stayed constant.
The only non-clock hardware registers that changed are \path{dc_fhout}, \path{rxInitAngle}, and \texttt{agcStatus}.
As shown in the histograms (\autoref{fig:dc_fhout}--\ref{fig:agcStatus}), these registers have very little variation. Moreover, they typically stay constant within one round and often
even constant over multiple rounds. \autoref{fig:s8_dcfhout} shows \path{dc_fhout} over time. In addition to the comparably
slow change in its value, it also shows a pattern.

The clock registers \path{dc_nbtc_clk} and \texttt{timer1value} change, but they are also not random. The Bluetooth clock, \path{dc_nbtc_clk},
is shared over-the-air. This is required to keep connections synchronous. In addition, the current Bluetooth clock value is used in substantial parts of
the protocol, i.e., as encryption algorithm input.
As shown in \autoref{fig:s8_clocks}, the hardware and Bluetooth clock are aligned. 
However, the hardware clock has a \SI{205}{times} higher granularity than the Bluetooth clock, meaning that even for a known
Bluetooth clock and a hardware clock value that leaked once, future hardware clock values still have a slight variance.

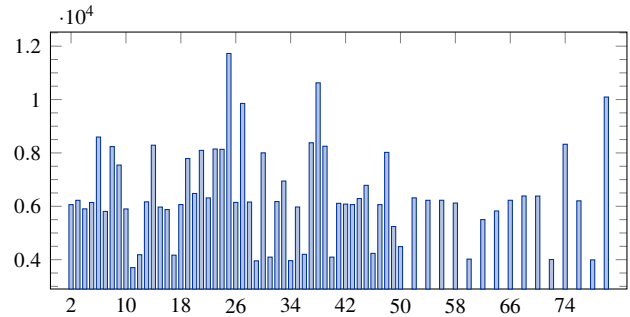
\begin{figure}[!b]
\begin{tikzpicture}[every node/.style={font=\footnotesize}]
\begin{axis}[ybar stacked, bar width=5.4pt, minor y tick num = 3, height=5cm, width=0.52\textwidth,
xticklabel style={rotate=90},
xticklabels={\texttt{0x0000}, \texttt{0x4cc2}, \texttt{0x4ccc}, \texttt{0x4cd8}, \texttt{0x4ce2}, \texttt{0x4cee}, \texttt{0x4cf8}, \texttt{0x4d04}, \texttt{0x4d0e}, \texttt{0x4d18}, \texttt{0x4d24}, \texttt{0x4d2e}, \texttt{0x4d3a}, \texttt{0x4d44}, \texttt{0x4d4e}, \texttt{0x4d5a}, \texttt{0x4d64}, \texttt{0x4d6e}, \texttt{0x4d7a}, \texttt{0x4d84}, \texttt{0x4d8e}, \texttt{0x4d9a}, \texttt{0x4df2}, \texttt{0x4dfc}, \texttt{0x4e08}, \texttt{0x4e12}, \texttt{0x4e1c}, \texttt{0x4e26}, \texttt{0x4e30}, \texttt{0x4e3c}, \texttt{0x4e46}, \texttt{0x4e50}, \texttt{0x4e5a}},
xtick={0, 1, 2, 3, 4, 5, 6, 7, 8, 9, 10, 11, 12, 13, 14, 15, 16, 17, 18, 19, 20, 21, 22, 23, 24, 25, 26, 27, 28, 29, 30, 31, 32},
xmin=-1, xmax=33.25]

\addplot[style = {darkblue, fill=darkblue!30}] coordinates {
        (0, 69632) (1, 135168) (2, 135168) (3, 118784) (4, 155648) (5, 114688) 
        (6, 126976) (7, 106496) (8, 122880) (9, 122880) (10, 126976) (11, 118784) 
        (12, 126976) (13, 143360) (14, 139264) (15, 131072) (16, 118784) (17, 131072) 
        (18, 110592) (19, 135168) (20, 118784) (21, 118784) (22, 135168) (23, 122880) 
        (24, 143360) (25, 122880) (26, 122880) (27, 139264) (28, 110592) (29, 122880) 
        (30, 98304) (31, 143360) (32, 106496) 
};
\end{axis}
\end{tikzpicture}
\caption{Histogram of values of \texttt{rxInitAngle} observed on a \emph{Samsung Galaxy S8}.}
\end{figure}

\clearpage

\subsection{PRNG Attacks}
\label{ssec:prngattack}

All variants, including the newest variant 5, access the \path{rbg_rand} function in the same way. An overview of the calling structure is shown in \autoref{fig:callgraph_to_rbg_rand_}.
How \path{rbg_rand} is used becomes relevant when attacking the \ac{PRNG}.
Attacks on the \ac{PRNG} can be divided into predicting (\autoref{sssec:predict}) and influencing (\autoref{sssec:influence}) its inputs as well as
extracting its current state (\autoref{sssec:extract}). Combining all of these provides the strongest attack vector.
However, an attacker can opt to not actively manipulate the \ac{PRNG} inputs, resulting in more bits to brute-force within
random numbers or to infer the precise internal \ac{PRNG} state.
We describe how an attacker can predict \ac{PRNG} outputs due to its weak implementation in \autoref{sssec:predictOutput}.

\subsubsection{Predicting PRNG Inputs}
\label{sssec:predict}

While some functions access \path{rbg_rand} directly, others use the wrapper \path{sha_get_128b_rand} that returns a \SI{16}{\byte} random number. This wrapper calls \path{rbg_rand} 16 times in a row
and applies the SHA-1 function. Note that SHA-1 is outdated and the Bluetooth specification recommends using SHA-256 within \acp{RNG}~\cite[p. 953]{bt52}. Since the timing of subsequent calls in a loop is predictable, \ac{PRNG} inputs that depend on time do not provide any additional entropy in this context.
Moreover, \path{dc_fhout}, \path{rxInitAngle}, and \texttt{agcStatus} stayed constant within one measurement round in most cases, and, thus, also do not add any entropy except from the value in the first round.

\subsubsection{Influencing PRNG Inputs}
\label{sssec:influence}

An attacker who has a crash-only over-the-air attack can reset the hardware clock to \texttt{0xFFFFFFFF}.
Since the Patchram slots in the \emph{Broadcom} and \emph{Cypress} chips are rare, only severe security issues can be patched.
Crash-only attacks, such as \emph{CVE-2019-6994}, often remain unpatched~\cite{classen2019inside}. The Bluetooth and hardware clock stay
correlated over time, meaning that an attacker does not necessarily need to crash the chip while the user is pairing a new device
or initiating an encrypted session, but any time before.

In addition to control and knowledge about the clock, a timed crash and following packet calling the \ac{PRNG} might also
set the current \ac{PRNG} status to a less random value.
The very first \ac{PRNG} round is initialized by copying memory from RAM (see line 11 in \autoref{lst:rbg_get_psrng_mac}), which this is likely filled
with predictable contents during chip initialization.

\begin{figure}[!b]
	\begin{center}
	\hspace*{-0.5em}\scalebox{0.76}{
		\tikzstyle{myarrow}=[->, >=latex', shorten >=1pt, thick]
		\tikzstyle{func_node}=[rectangle, draw=black, rounded corners, fill=white, text centered, anchor=north, text=black, font=\ttfamily \footnotesize]
		
		\begin{tikzpicture}[node distance=2mm and 4mm, auto]
		\node[minimum height=2em, minimum width=12.5em, text height=1em, align=center, func_node]
		(ape_rand_other) {\dots};
		
		\node[minimum height=2em, minimum width=12.5em, text height=1em, align=center, func_node,below= of ape_rand_other]
		(sp_generate_PPKeyPair) {sp\_generate\_PPKeyPair};
		
		\node[minimum height=2em, minimum width=12.5em, text height=1em, align=center, func_node, yshift=-3mm, below= of sp_generate_PPKeyPair]
		(ulp_rand_other) {\dots};
		
		\node[minimum height=2em, minimum width=12.5em, text height=1em, align=center, func_node,below= of ulp_rand_other, fill=darkred!10, draw=darkred, thick]
		(smulp_genIV) {smulp\_genIV};
		
		\node[minimum height=2em, minimum width=12.5em, text height=1em, align=center, func_node, yshift=-3mm, below= of smulp_genIV]
		(sha_others) {\dots};
		
		\node[minimum height=2em, minimum width=12.5em, text height=1em, align=center, func_node,below= of sha_others]
		(lm_HandleHCIMasterLinkKey) {lm\_HandleHCIMasterLinkKey};
		
		\node[minimum height=2em, minimum width=12.5em, text height=1em, align=center, func_node,below= of lm_HandleHCIMasterLinkKey]
		(ap_action_txStartEncryptReq) {ap\_action\_txStartEncryptReq};
		
		\node[minimum height=2em, minimum width=12.5em, text height=1em, align=center, func_node,below= of ap_action_txStartEncryptReq]
		(lm_HandleHciPinCodeReqReply) {\_ape\_action\_txCombKey}; 
		
		\node[minimum height=2em, minimum width=12.5em, text height=1em, align=center, yshift= 4mm, func_node,right= of sp_generate_PPKeyPair]
		(ape_rand) {ape\_rand};
		
		\node[minimum height=2em, minimum width=12.5em, text height=1em, align=center, yshift= 4mm, func_node, right= of smulp_genIV, fill=darkred!10, draw=darkred, thick]
		(ulp_rand) {ulp\_rand};
		
		\node[minimum height=2em, minimum width=12.5em, text height=1em, align=center, yshift= 4mm, func_node,right= of ap_action_txStartEncryptReq, fill=darkblue!10, draw=darkblue, thick]
		(sha_get_128b_rand) {sha\_get\_128b\_rand};
		
		\node[minimum height=2em, minimum width=12.5em, text height=1em, align=center, yshift= -4mm, func_node,below= of sha_get_128b_rand]
		(lculp_createAccessAddress) {lculp\_createAccessAddress};
		
		\node[minimum height=2em, minimum width=12.5em, text height=1em, align=center, yshift= -3mm, func_node,below = of lculp_createAccessAddress]
		(lculp_initSetConnDefaultParam) {lculp\_initSetConnDefaultParam};
		
		\node[minimum height=2em, minimum width=12.5em, text height=1em, align=center, func_node, yshift=3mm, above= of ape_rand]
		(rbg_rand_other) {\dots};

		\node[minimum height=2em, minimum width=4em, text height=1em, align=center, func_node, yshift=-1.6cm, right= of ulp_rand, fill=darkblue!10, draw=darkblue, thick]
		(rbg_rand) {rbg\_rand};

		\draw[myarrow] (ape_rand_other.east) -- (ape_rand.west);
		\draw[myarrow] (sp_generate_PPKeyPair.east) -- (ape_rand.west);
		
		\draw[myarrow] (ulp_rand_other.east) -- (ulp_rand.west);
		\draw[myarrow] (smulp_genIV.east) -- (ulp_rand.west);
		
		\draw[myarrow] (sha_others.east) -- (sha_get_128b_rand.west);
		\draw[myarrow] (lm_HandleHCIMasterLinkKey.east) -- (sha_get_128b_rand.west);
		\draw[myarrow] (ap_action_txStartEncryptReq.east) -- (sha_get_128b_rand.west);
		\draw[myarrow] (lm_HandleHciPinCodeReqReply.east) -- (sha_get_128b_rand.west);
		
		\draw[myarrow] (rbg_rand_other.east) -- (rbg_rand.west);
		\draw[myarrow] (ape_rand.east) -- (rbg_rand.west);
		\draw[myarrow] (ulp_rand.east) -- (rbg_rand.west);
		\draw[myarrow] (sha_get_128b_rand.east) -- (rbg_rand.west);
		\draw[myarrow] (lculp_createAccessAddress.east) -- (rbg_rand.west);
		\draw[myarrow] (lculp_initSetConnDefaultParam.east) -- (rbg_rand.west);
		
		\draw[myarrow] (lculp_initSetConnDefaultParam.north) -- (lculp_createAccessAddress.south);

		\end{tikzpicture}
		}
		\caption{Call graph to \texttt{rbg\_rand} on the \emph{CYW20735B1} evaluation board.}
		\label{fig:callgraph_to_rbg_rand_}
	\end{center}
\end{figure}
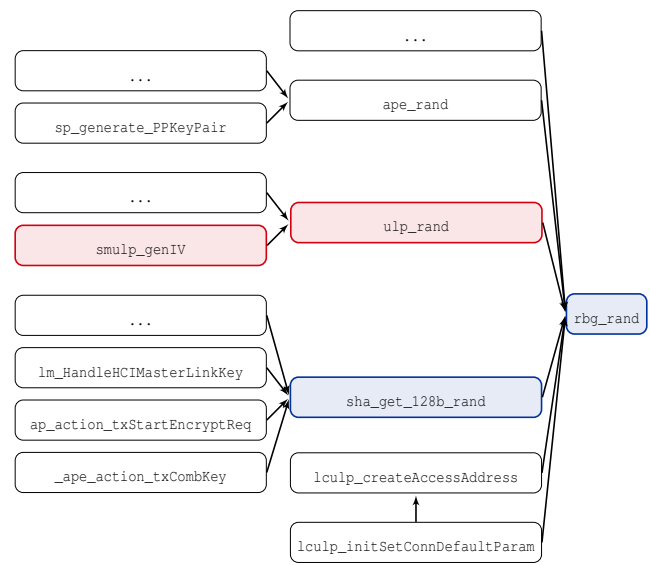

\subsubsection{Extracting PRNG Outputs}
\label{sssec:extract}

An attacker needs to know the current \ac{PRNG} state including the previous random value \path{Rand}
to get full knowledge about future \ac{PRNG} values.
Assuming that the firmware already accesses the \ac{PRNG} a few times during initialization, even a 
chip reset might leave the attacker with some variability. Moreover, an attacker might not be able
to reset the chip and, thus, does not have any knowledge about the internal \ac{PRNG} state and needs to brute-force it.

All \ac{BLE}-related functions access \path{ulp_rand}, which in turn calls \path{rbg_rand}.
Thus, the \path{LE_Rand} \ac{HCI} command allows direct access to the current \ac{RNG} state.

However, an attacker typically acts over-the-air and does not have access to the host.
The current state of the \ac{RNG} can be leaked over-the-air as follows.
An attacker can send an \path{LL_ENC_REQ}~\cite[p. 2898]{bt52}, which is answered with an \path{LL_ENC_RSP}~\cite[p. 2899]{bt52}.
The \path{LL_ENC_RSP} by the device under attack contains the fields \path{SKDs} and \path{IVs}. Within the firmware, these are generated by the functions \path{smulp_genSKD} 
and \path{smulp_genIV}, which both call \path{ulp_rand} that directly accesses the \ac{PRNG}.
This implementation is the same for the \emph{CYW20735B1} evaluation board, which is shown in the call graph in \autoref{fig:callgraph_to_rbg_rand_}, and the \emph{Samsung Galaxy S8}.

Regarding the \emph{Samsung Galaxy S8}, the \emph{Android} implementation mentioned in \autoref{ssec:androidle}
is vulnerable to passive \ac{MITM} attacks. An attacker does not need to establish a connection and send 
an \path{LL_ENC_REQ} \ac{PDU}---\emph{Android} transmits the \ac{BLE} \ac{RNG} state within various packet types.


An attacker who inferred the current \ac{PRNG} state and has knowledge or control about the hardware clock
can calculate upcoming values provided by the \ac{PRNG}.

\todo{maybe be explicit and say that BLE and Classic use the same RNG and so attacking either of them is the same}

\todo{crc32 backwards calculation? or at least we have so few inputs that we can build some tables. we know which value might follow after which.}


\subsubsection{Predicting PRNG Outputs}
\label{sssec:predictOutput}

We will now describe how to predict randomness generated by the \ac{PRNG} based on previous outputs. For each \SI{4}{\byte} randomness, the \ac{PRNG} uses the previous output concatenated with the internal clock, the Bluetooth clock, and a number of registers. Then, it hashes them using a \acf{CRC}. Note that the input state consists of \SI{32}{\byte}, therefore, it is infeasible to determine the full state based on outputs of the \ac{PRNG}, even if we know that some of the values are not uniformly random. However, an attacker can exploit the fact that CRC32 is an affine function, using an initialization value of $IV=$\texttt{0xFFFFFFFF}. For two inputs $a$ and $b$, \[\text{CRC32}(a)\oplus\text{CRC32}(b)=\text{CRC32}(a\oplus b)\oplus\text{CRC32}(IV).\]

\noindent
Due to this affinity of CRC32, if we know the two previous outputs of the \ac{PRNG}, we only need to guess the difference in the inputs to the \ac{PRNG}.
For the Bluetooth clock, we know its value as it is included in every transmission, and for the system clock, only the lowest bits will change with high probability. Further, as shown in the previous sections, the entropy in the other registers is quite low. Let $out_0$ and $out_1$ be the two previous outputs, then
\[out_2 = out_1 \oplus \text{CRC32}( C || R || P ) \oplus \text{CRC32}(IV),\]
where $C$ is the guessed bit difference in both clocks, $R$ is the guessed bit difference in the registers, and $P$, the bit difference to the previous output field, is given by
\[ P = \text{CRC32}(out_0) \oplus \text{CRC32}(out_1) \oplus \text{CRC32}(IV).\]

We estimate the difference entropy of the registers to be less than \SI{18}{\bit}. If we assume we have a synchronization between the Bluetooth clock and the hardware clock as described in \autoref{sssec:influence}, we need \SI{8}{\bit} for the hardware clock, resulting in a total of \SI{26}{\bit}. Note that further calls to the \ac{PRNG} afterward are deterministic with high probability.

For example, for Numeric Comparison, we start by choosing a  \SI{128}{\bit} random number $Nb$, which is generated by \num{16} successive calls to the \ac{PRNG}. As we have shown in \autoref{ssec:numericComparison}, the knowledge of $Nb$ would enable an \ac{MITM} attack on the pairing process. 
Further, due to the deterministic commitment, we can check as an \ac{MITM} attacker whether we found the correct $Nb$. A brute force attack to check all possible outputs of the \ac{PRNG} would take about \SI{5}{\minute} on a single CPU core. As this is trivially parallelizable, this is a realistic attack given enough computing power. 
Similar efficiency is to be expected in the case of generated private keys using randomness from the \ac{PRNG}, as discussed in \autoref{ssec:androidle}.


%% file: sections/discussion.tex

\section{Discussion}
\label{sec:discussion}

Initially, we only measured the \ac{PRNG} fallback on the \emph{Cypress} evaluation boards and the \emph{Google Nexus 5}.
However, we found that the \ac{PRNG} was not accessed during regular usage on those devices. Nonetheless,
we suspected it being used in other chips due to the observed code changes, and, thus, reported it to \emph{Broadcom},
\emph{Cypress}, \emph{Apple}, \emph{Google}, and \emph{Samsung} on January 12, 2020. We also informed
the maintainer of \emph{BTstack}~\cite{btstack} on the same date, because we observed that they were
excessively using the \ac{HCI} \path{LE_Rand} function during initialization of the Bluetooth stack
for key generation. After a discussion with the \emph{BTstack} maintainer, we decided to test if the
\ac{PRNG} was accessed on the \emph{Raspberry Pi 3/3+} when constantly calling \path{LE_Rand} for
more than a day, but luckily, the results were negative.

After this first round of responsible disclosure, \emph{Broadcom} claimed that the \ac{PRNG} fallback
would not be used on any of their devices. This is when we started analyzing \num{20} different firmware versions. 
We informed \emph{Broadcom}, \emph{Samsung}, and \emph{Google} about the missing \ac{HRNG} in the \emph{Samsung Galaxy S8}
on February 1, and also updated the others about the possibility that there are indeed chips without \ac{HRNG}.
\emph{Google} closed the issue as \emph{Won't Fix (Infeasible)} on February 4, because it is up to \emph{Broadcom} to fix the firmware
and none of their products is affected.
A patch for the \emph{Samsung Galaxy S8} as well as its variants \emph{S8+} and \emph{Note 8} was released in May.

There are a few \emph{iPhone} models that we did not test, and the \emph{iOS 13.5} release contains a patch for
a model that also missed a proper \ac{RNG}. Since a test requires a jailbroken \emph{iPhone} and we do not have all \emph{iPhone}
models available for testing, we could not identify the precise model and also could not perform any measurements. However, we performed tests
for the registers used by the fallback on other devices and they provide similar properties as those on the \emph{Samsung Galaxy S8}.

As \emph{Samsung} is using \emph{Broadcom} as well as \emph{Qualcomm} chips on their smartphones, depending on
the market and device model, they asked if they could forward our report. Since we did not
find anything on \emph{Qualcomm} chips ourselves and \emph{Qualcomm} has an \ac{NDA} with \emph{Samsung}
possibly allowing them to include confidential information, we asked \emph{Samsung} to forward the 
report and exclude us if needed. We received the following answer on March 3:

\begin{displayquote}
\emph{``We haven't found any indicators that our Bluetooth implementations are affected by the PRNG issue after internally discussing the research results that were shared with us.''} \newline \vspace*{-0.5em}\hspace*{14.5em}QPSIIR-1329
\end{displayquote}


What remains unclear is how the \emph{Samsung Galaxy S8} happened to miss an \ac{HRNG} in the first place.
The code for accessing the \ac{HRNG} is already missing in the firmware, indicating that this issue was most
likely known to developers during compilation---if not optimized and automated by another process.
Yet, the \emph{BCM4347B0} chip made it into such a popular smartphone.


%% file: sections/conclusion.tex

\section{Conclusion}
\label{sec:conclusion}

The development over a decade within the \emph{Broadcom} and \emph{Cypress} chips indicates that the \ac{RNG}
is indeed a central component. The firmware history shows how the \ac{PRNG} fallback was first improved and then
removed completely. However, removing it might also be harmful. If the \ac{PRNG} fallback had
been missing in the \emph{Samsung Galaxy S8}, it might have simply returned static values.

Another interesting detail in the firmware history is that \emph{Cypress} independently developed the \ac{PRNG}
code after acquiring the \ac{IoT} branch of \emph{Broadcom} in 2016~\cite{cypressbroadcom}.
This indicates that also the patching process might differ depending on which company is assigned to which chip,
even though the firmware has similar issues.

Overall, the \ac{RNG} provided by a Bluetooth chip remains a black box without intensive analysis.
The Bluetooth specification requires an \ac{RNG} that passes tests such as the \emph{Dieharder} test suite.
This requirement leads to the potentially false assumption that a Bluetooth chip's \ac{RNG} can be trusted for
security-relevant operations, including the \ac{HCI} \path{LE_Rand} command exposing the \ac{RNG}
to energy-constrained embedded devices and the \emph{Android} Bluetooth stack.
Our \ac{RNG} testing scripts are publicly available, allowing benchmarks of future or not yet
tested \emph{Broadcom} and \emph{Cypress} chips as well as porting these concepts to chips of other
manufacturers.

We tested many Bluetooth chips on our own. However, it would be helpful if manufacturers made details
about the \ac{RNG} openly accessible. During our research, we only found that the non-Bluetooth \emph{Broadcom}
main \ac{SoC} on the \emph{Raspberry Pi} might include an \ac{HRNG} that is based on thermal noise~\cite{bcmrng}. More transparency, i.e.,
statements whether a chip includes an \ac{HRNG} or a \ac{PRNG} and which type exactly, would be helpful when
making design decisions on embedded and mobile devices.
